\documentclass[a4paper,twoside,reqno,11pt]{article}
\usepackage{amssymb,amsmath,amsthm}
\usepackage{graphicx}
\usepackage{times}
\usepackage{color}
\usepackage{natbib}
\usepackage{float}
\leftmargini=\baselineskip
\newtheoremstyle{localthm}
	{7pt} 
	{7pt} 
	{\sl} 
	{} 
	{\bf} 
	{{\rm.}} 
	{.7em} 
	{} 

\theoremstyle{localthm}

\newtheoremstyle{localrem}
	{5pt} 
	{5pt} 
	{\rm} 
	{} 
	{\bf} 
	{{\rm.}} 
	{.7em} 
	{} 

\theoremstyle{localrem}

\def\rss{\mathrm{rss}}
\def\RSS{\mathrm{RSS}}

\newcommand{\R}{\mathbb{R}}

\def\rss{\mathrm{rss}}
\def\RSS{\mathrm{RSS}}

\def\bs{\boldsymbol}

\def\x{\bs{x}}
\def\X{\bs{X}}
\def\y{\bs{y}}
\def\Y{\bs{Y}}

\def\Z{\bs{Z}}
\def\P{\bs{P}}

\def\hat{\widehat}
\def\tilde{\widetilde}

\begin{document}
\addtolength{\baselineskip}{0.4\baselineskip}

\title{Linear Regression, Covariate Selection and the Failure of Modelling.}
\author{Laurie Davies\\
University of Duisburg-Essen\\
Emails:laurie.davies@uni-due.de}
\date{\today}
\maketitle
\begin{abstract}
It is argued that all model based approaches to the selection of covariates in linear regression have failed. This applies to frequentist approaches based on P-values and to Bayesian approaches although for different reasons. In the first part of the paper 13 model based procedures are compared to the model-free Gaussian covariate procedure in terms of the covariates selected and the time required. The comparison  is based on seven data sets and three simulations. There is nothing special about these data sets which are often used as examples in the literature. All the model based procedures failed.

In the second part of the paper it is argued that the cause of this failure is the very use of a model. If the model involves all the available covariates standard P-values can be used. The use of P-values in this situation is quite straightforward. As soon as the model specifies only some unknown subset of the covariates the problem  being to identify this subset the situation changes radically. There are many P-values, they are dependent and most of them are invalid. The P-value based approach collapses.  The Bayesian paradigm also assumes a correct model but although there are no conceptual problems with a large number of covariates there is a considerable overhead causing computational and allocation problems even for moderately sized data sets. 

The Gaussian covariate procedure is based on P-values which are defined as the probability that a random Gaussian covariate is better than the covariate being considered. These P-values are exact and valid whatever the situation. The allocation requirements and the algorithmic complexity are both linear in the size of the data making the procedure capable of handling large data sets. It outperforms all the other procedures in every respect.
\end{abstract}
{\it Keywords}. linear regression, covariate selection, models, real data

\section{Introduction} \label{sec:intro}
The claim that model based procedures have failed will be made in Section 2. It is based on  a comparison of the Gaussian covariate method  \cite{DAVDUEM22} with the following 13 selection procedures: lasso (\cite{TIB96}), knockoff (\cite{CAFAJALV2018}), scaled sparse linear regression (\cite{SUNZHA12}), SIS (Sure Independence Screening) (\cite{FANLV08}), desparsified lasso (\cite{ZHAZHA14}), stability selection (\cite{MEIBUE10}), ridge regression (\cite{BUEH13}), multiple splitting (\cite{WASSROED09}), EMVS (Expectation-Maximization Approach to Bayesian Variable Selection) (\cite{ROCGEO14}) and   Spike and Slab Regression (\cite{SCO21}), Threshold Adaptive Validation (\cite{LFL21}), graphical lasso (\cite{FHT08,FRHATI19})  and huge (High-Dimensional Undirected Graph Estimation) (\cite{JFLRLWLZ21}. The latter three construct dependency graphs. A short description of the Gaussian covariate method which is sufficient for this paper is given in the Appendix.

The comparison is based on three simulations and the following seven data sets: riboflavin \cite{BUEKALMEI14}, leukemia \cite{GOLETAL99}, lymphoma \cite{ALI00} and \cite{DETBUH02}, osteoarthritis \cite{COXBATT17}, the Boston Housing data set \cite{HARRUB78}, sunspot data \cite{SIDB} and the American Business Cycle data  \cite{ABC86}.  Their dimensions $(n,q)$ are respectively (71,4088), (72,3571), (62,4026),(129 48802), (506,176358), (3253, 3252) and (240,22). The large number of covariates, 176358, for the Boston Housing data is that these are all interactions of degree at most eight of the original 13 covariates. 

The reasons for this failure are discussed in Section 3. Statistical models are often successful so a case where they fail is of particular interest, the issues involved are not limited to this particular failure.

\section{Comparing procedures}
Two Gaussian procedures {\it f1st} and {\it f3st} with $p0=0.01$ and $kmn=10$ (see \cite{DAVDUEM22} and the Appendix), will be compared with the model based procedures. All these are used with their default values, for example $B=100$ in  {\it multi.split}, with the exceptions of {\it stability} which requires a value for $EV$ set here to $EV=1$ and {\it lm.spike} which requires a value for $niter$ set here to $niter=1000$. 

\subsection{Riboflavin and leukemia simulations} \label{sec:ribo_sim}
The simplest linear model for formulating the problem of covariate selection is
\begin{equation} \label{equ:model}
\Y=\sum_{\x_i\in {\mathcal S}_0}\beta_i\x_i+\sigma\bs{\varepsilon}
\end{equation}
where $\bs{\varepsilon}$ is standard Gaussian nose and ${\mathcal S}_0$ is a subset of the $q$ covariates, the so called active set, with $\beta _i \ne 0$ for $\x_i\in {\mathcal S}_0$. The problem is to determine ${\mathcal S}_0$ and to construct confidence intervals and P-values for the corresponding values of the $\beta_i$ ( see \cite{DEBUEMEME15}). Having postulated the model (\ref{equ:model}) the statistician will now work within it, that is, treating the model as true. 

The following simulations based on the model (\ref{equ:model}) were performed but using the 4088 covariates of the riboflavin data and the 3571 covariates of the leukemia data . The covariates were first standardized to have mean zero and variance one. Four covariates $\{\x_{i_1},\x_{i_2},\x_{i_3},\x_{i_4}\}$ are chosen at random and the dependent variable given by
\[\y=20\sum_{j=1}^4 \x_{i_j} +\bs{\varepsilon}\]
where $\bs{\varepsilon}$ is standard Gaussian noise. The reason for taking just four covariates is that the simplest Gaussian covariate procedure applied to the riboflavin data results in just the four covariates $\x_{4003}, \x_{2564}, \x_{73}, \x_{2034}$. In the case of the leukemia data six covariates were chosen at random.

All ten of the model based procedures were considered using the functions  {\it cv.glmnet}  for lasso (\cite{FRHATINATASNQI21}), {\it knockoff.filter} for knockoff (\cite{BCJPS20}), {\it scalreg} for scaled sparse linear regression (\cite{SUN19}) , {\it SIS} for Sure Independence Screening (\cite{FFSS20}), {\it stability} for stability selection, {\it multi.split} for multiple splitting , {\it ridge.proj} for ridge regression, {\it lasso.proj} for desparsifed regression (the last four available from \cite{MDMMB21}),  {\it EMVS} for EMVS variable selection (\cite{ROCMOR21}) and {\it lm.spike} for BoomSpikeSlab (\cite{SCO21}). Two Gaussian covariate functions were used, {\it f1st} and {\it f3st} with $kmn=10$ and with two different values of the parameter $m=1,2$ of {\it f3st}.

Two, {\it knockoff} and {\it desparse.lasso}, required more than one hour for a single simulation and were killed after this time. {\it multi.split} often halted with an error, this occurred  after 70 simulations in Table~\ref{tab:ribo_sim} and 38 simulations in Table~\ref{tab:leuk_sim}. {\it BoomSpikeSlab} in conjunction with the operating system Leap 15.1 occasionally shut the laptop down, this happened after 72 simulations with the riboflavin data. {\it ridge} is very slow and was killed after 18 simulations.  For the remaining procedures 100 simulations were performed. 

\begin{table}[h]
\begin{center}
\begin{tabular}{rcccc}
\multicolumn{5}{c}{Riboflavin: 100 simulations; (*) 70, (**) 72, (***) 18 simulations }\\
\hline\\
method&$fp$&$fn$&\% correct&time\\
{\it f1st}&0.77&0.72&75&1 (0.026)\\
{\it f3st,m=1}& 0.18& 0.17&93& 5\\
{\it f3st,m=2}& 0.07&0.05&98&24\\
{\it lasso}&25.0& 0.07& 0&19\\
{\it scalreg}& 16.3&1.08& 0&85\\
{\it SIS} & 13.5& 2.45&3&150\\
{\it stability}& 0.24& 2.16&8&96\\
{\it multi.split(*)}&0.23&1.59&36&1570\\
{\it BoomSpikeSlab(**)}& 0.33& 0.42&87&1540\\
{\it EMVS} & 0.00& 4.00& 0&27\\
{\it knockoff}& ?& ?&?&$>$150000\\
{\it desparse.lasso}& ?& ?&?&$>$150000\\
{\it ridge(***)}& 0.00&4.00&0.00&10000\\
\end{tabular}
\caption{Columns 2-4 give the average number of false positives, the average number of false negatives and the \% of correct selections. Column 5 gives time compared with Gaussian covariates which required on average 0.026 seconds per simulation. \label{tab:ribo_sim}}
\end{center}
\end{table}

\begin{table}[h]
\begin{center}
\begin{tabular}{rcccc}
\multicolumn{5}{c}{Leukemia: 100 simulations, (*) 38, (**) 18 simulations}\\
\hline\\
method&$fp$&$fn$&\% correct&time\\
{\it f1st}&0.43&1.21&71&1 ( 0.025)\\
{\it f3st,m=1}& 0.26& 0.68&79& 7\\
{\it f3st,m=2}& 0.29& 0.69&81&31\\
{\it lasso}&19.8& 0.11& 0&18\\
{\it scalreg}&10.8&1.55& 0&144\\
{\it SIS} & 11.6& 2.35&6&130\\
{\it stability}& 0.04&4.73&1&94\\
{\it multi.split (*)}&0.10&4.7&3&1500\\
{\it BoomSpikeSlab}& 0.50&3.15&38&1340\\
{\it EMVS} & 0.00& 6.00& 0&19\\
{\it knockoff}& ?& ?&?&$>$150000\\
{\it desparse.lasso}& ?& ?&?&$>$150000\\
{\it ridge(**)}& 0.00&6.00&0.00&7500\\
\end{tabular}
\caption{Columns as for Table~\ref{tab:ribo_sim}. \label{tab:leuk_sim}}
\end{center}
\end{table}

It is clear from Tables~\ref{tab:ribo_sim}  and ~\ref{tab:leuk_sim} that the Gaussian procedures are overwhelmingly superior. This is in spite of the fact that the data were generated under the model (\ref{equ:model}). In terms of covariate selection {\it BoomSpikeSlab} was the best of the model based procedures but worse than {\it f3st}  with $kmn=10$ and $m=1$, it was also between 190 and 300 times slower.  When applied to the osteoarthritis and the Boston Housing datasets (see Tables~\ref{tab:osteoarthritis} and \ref{tab:boston} below) {\it BoomSpikeSlab} fails completely.
\clearpage
 \subsection{The riboflavin, the leukemia, the lymphoma, the osteoarthritis, the Boston Housing, the sunspot and the American Business Cycle data sets}
 Table~\ref{tab:gaucov_results} gives the detailed results for the first five data sets using the default version {\it f1st} of the stepwise procedure with $kmn=10$. The first rows give in order the data set followed by the selected covariates. The second rows gives the time in seconds and the sum of squared residuals followed by the Gaussian P-values of the selected covariates.
\begin{table}[ht]
\begin{center}
\begin{tabular}{rccccccc}
\multicolumn{8}{c}{Gaussian covariate results for the first five data sets}\\
\hline
riboflavin&4003&2564&73&2034&*&*&*\\
 0.02, 8.448&1.97e-14& 2.81e-13& 4.1e-09& 2.84e-05&*&*&*\\
\hline
leukemia& 1182&1219&2888&183&3038&2558&*\\
0.181, 0.931&2.7e-07&3.23e-07&4.90e-05& 1.47e-04&4.47e-03&8.00e-03&*\\
\hline
lymphoma& 2805&3727&714&2036&3201&1638&2057\\
0.024, 0.962&1.86e-06&2.30e-15&3.75e-07& 4.29e-09&8.19e04&5.55e-03&9.96e-03\\
\hline
Osteoarthritis&1149&31848&33321&*&*&*&*\\
0.13, 6.767&3.80e-18&1.54e-05&6.25e-03&*&*&*&*\\
\hline
Boston inter.&7222&164793&106761&81454&29154&170455&67241\\
4.59, 6129.2&1.95e-143&2.15e-10&4.65e-07&8.13e-13&1.62e-16 &1.18e-52&1.99e-04\\
\end{tabular}
\caption{The Gaussian covariate results for the data sets  riboflavin, osteoarthritis and Boston Housing \label{tab:gaucov_results}}
\end{center}
\end{table}

Table~\ref{tab:gaucov_results} gives only one approximation for each data sets. More can be obtained by using either {\it f2st} or {\it f3st}, the latter usually giving better results. Applying {\it f3st} to the riboflavin data with $kmn=15, m=5$ results in 129 approximations. The first 10 are given in Table~\ref{tab:f3st} in order of the sum of squared residuals. The approximation in Table~\ref{tab:gaucov_results} is 32nd on the list.

\begin{table}[h]
\begin{center}
\begin{tabular}{cccccccccc}
$\bs{ss}$&\multicolumn{9}{c}{Riboflavin: Included covariates}\\
\hline
 3.72& 4004& 2564&   73&  315& 2936 & 997&  991& 1661&  3255\\
4.23& 4004& 2564 &  73 & 315& 2936&  997& 1661& 2048&*\\
  4.87& 4004& 2564 &144& 1131& 3138& 2186& 1771&*&*\\
  5.43& 1279& 4004& 2564&   73& 1131& 2140&    *&*&*\\
  5.47& 4003& 2564 &  69&1425&  413& 2484& 1194&*&*\\
  5.53&1279& 4004& 2564&  144 &1131& 2140&   *&*&*\\
  5.57& 1279& 2564 &  73& 1131& 2140& 4005&   *&*&*\\
  5.71& 1278& 4004 &2564 &  73 &1131& 2140 &   *&*&*\\
5.71&4004 &1304& 2564&   73& 1131& 2140&*&*&*\\
  5.80& 1278& 4004& 2564&  144& 1131& 2140&    *&*&*\\
\end{tabular}
\caption{The first ten of the 129 approximations given by {\it f3st} with $kmn=15$ and $m=5$ in order of the sum of squared residuals $\bs{ss}$. \label{tab:f3st}}
\end{center}
\end{table}

In the Tables~\ref{tab:riboflavin}, \ref{tab:leukemia},  \ref{tab:osteoarthritis},  \ref{tab:boston} and  \ref {tab:sunspot_lag} the first column gives the method and the second the numbers $k$ of covariates selected and the number $fp$ of selected covariates which had Gaussian P-values exceeding 0.99.   The third column indicates whether the procedure returns P-values. The fourth gives the sum of squared residuals. The fifth gives the factor $t/t0$ where $t$ is the time in seconds of the procedure and $t0$ is the time for the Gaussian covariate method {\it f1st} with $kmn=10$.  The results show that all ten modelled based procedures failed for all seven data sets.  Note that {\it f1st}, {\it f3st}, {\it scalreg} and {\it EMVS} are deterministic and give the same result when repeated. The others make use of random choices and if repeated may give somewhat different results. 

When applied to the riboflavin data Table~\ref{tab:riboflavin} {\it knockoff} was killed after two hours and had at  this point selected zero covariates. For this reason it was not applied to the remaining data sets. Similarly  {\it desparsified} was very slow and was applied only to the riboflavin, leukemia and the lagged sunspot and American Business Cycle data sets. The osteoarthritis data set was too large for {\it EMVS} and {\it BoonSpikeSlab}. This left only five model based procedures for the Boston Housing data Table~\ref{tab:boston}. Three failed with errors presumably because of the size of the data. {\it Lasso} returned 63 covariates 61 of which had Gaussian P-values of 1. In further  applications of {\it lasso} it returned between four and over 100 covariates. {\it Stability} returned just three covariates and took over 11 minutes to do so.

\begin{table} [ht]
\begin{center}
\begin{tabular}{rcccc}
\multicolumn{5}{c}{riboflavin (71,4088)}\\
\hline\\
method& $k,fp$&P-values&$\bs{ss}$&time\\
\hline\\
{\it f1st}&4,0&yes& 8.45&1 (0.024)\\
{\it f3st,m=1}&6,0&yes&6.21&4\\
{\it f3st,m=2}&6,0&yes&5.43&22\\
{\it lasso}&32,30&no&2.05&25.7\\
{\it knockoff}&0,0&no&*&$>$7e+05 (killed)\\
{\it scalreg}&9,6&no&10.62&28.7\\
{\it SIS}&4,0&no&11.49&89\\
{\it desparsified lasso}&0,0&yes&*&130012\\
{\it stability}&0,0&no&*&103\\
{\it ridge.proj}&0,0&yes&*&12248\\
{\it multi.split}&4,2&yes&17.45&1421\\
{\it EMVS}&0,0&no&*&22\\
{\it BoomSpikeSlab}&(5,2)&no&16.92&2290\\
\end{tabular}
\caption{The results for the riboflavin data. \label{tab:riboflavin}}
\end{center}
\end{table}

\begin{table}[ht]
\begin{center}
\begin{tabular}{rcccc}
\multicolumn{5}{c}{leukemia $(n,q)=(72,3571)$}\\
\hline\\
method& $k,fp$&P-values&$\bs{ss}$&time\\
\hline\\
{\it f1st}&6,0&yes&0.93&1 (0.0212)\\
{\it f3st, m=1}&6,0&yes&0.49&7\\
{\it f3st, m=2}&6,0&yes&0.49&30\\
{\it lasso}&17,14&no&0.661&20\\
{\it scalreg}&14,10&no&1.08&14\\
{\it SIS}&3,0&no&1.89&85\\
{\it desparsified lasso}&4,1&yes&1.56&104670\\
{\it stability}&1,0&no&5.43&103\\
{\it ridge.proj}&0,0&yes&*&11515\\
{\it multi.split}&1,0&yes&4.42&1420\\
{\it EMVS}&0,0&no&*&20\\
{\it BoomSpikeSlab}&(3,3)&no&13.73&1600\\
\end{tabular}
\caption{The results for the leukemia data. \label{tab:leukemia}}
\end{center}
\end{table}
\begin{table}[hb]
\begin{center}
\begin{tabular}{rcccc}
\multicolumn{5}{c}{Lymphoma data  (62, 4026)}\\
\hline\\
method&$k,fp$&P-values&$\bs{ss}$&time\\
\hline\\
{\it f1st}&7,0&yes&0.96&1 (0.024)\\
{\it f3st,m=1}&7,0&yes&0.96&8\\
{\it lasso}&43,42&no&0.058&21\\
{\it scalreg}&42,41&no&0.083&32\\
{\it SIS}&5,0&no&1.75&430\\
{\it desparsified lasso}&5,0&yes&1.62&111180\\
{\it stability}&0,0&no&*&90\\
{\it ridge.proj}&0,0&yes&*&11700\\
{\it multi.split}&0,0&yes&*&1540\\
{\it EMVS}&0,0&no&*&20\\
{\it BoomSpikeSlab}&(6,4)&no&6.82&1650\\
\end{tabular}
\caption{The results for the  lymphoma data. \label{tab:lymphoma}}
\end{center}
\end{table}

\begin{table}[hb]
\begin{center}
\begin{tabular}{rcccc}
\multicolumn{5}{c}{Osteoarthritis (129,48802)}\\
\hline\\
method&$k,fp$&P-values&$\bs{ss}$&time\\
\hline\\
{\it f1st}&3,0&yes& 6.77&1 (0.29)\\
{\it f3st,m=1}&3,0&yes& 6.23&4\\
{\it f3st,m=2}&4,0&yes&5.02&16\\
{\it lasso}&66,64&no&0.35&24.5\\
{\it scalreg}&5,3&no&11.14&45\\
{\it SIS}&0,0&no&*&80\\
{\it stability}&1,0&no&13.3&116\\
{\it multi.split}&3,1&yes&8.35&1470\\
{\it EMVS}&0,0&no&*&bad alloc.\\
{\it BoomSpikeSlab}&0,0&no&*&17.7 Gb\\
\end{tabular}
\caption{The results for the osteoarthritis data. \label{tab:osteoarthritis}}
\end{center}
\end{table}

\begin{table}[ht]
\begin{center}
\begin{tabular}{rcccc}
\multicolumn{5}{c}{Boston data with interactions (506,176358)}\\
\hline\\
method&$k,fp$&P-values&$\bs{ss}$&time\\
\hline\\
{\it f1st}&7,0&yes&6130&1 (4.42)\\
{\it f3st,m=1}&9,0&yes&5589&8\\
{\it f3st,m=3}&10,0&yes&5557&68\\
{\it lasso}&4,2&no&9950&30\\
{\it SIS}&0,0&no&*&max.iter\\
{\it scalreg}&0,0&no&*&self killed\\
{\it stability}&4,1&no&14375&160\\
{\it multi.split}&0,0&yes&*&incorrect no. of P-values\\
\end{tabular}
\caption{The results for the Boston Housing data. \label{tab:boston}}
\end{center}
\end{table}

The results for the sunspot data are given in Tables~\ref{tab:sunspot} and \ref{tab:sunspot_lag}. The covariates in Table~\ref{tab:sunspot} are $\x_{2j}=\sin(\pi jt)$ and $\x_{2j-1}=\cos(\pi jt)$ for $j=1,\ldots,1626$ with $t=(1:3253)/3253$. 

Table~\ref{tab:sunspot_lag} the covariates are the first 500 lags of the number of sunspots. The lags selected by {\it f1st} were 1, 2 ,  4,   6,   9,  27 and 117. We note that the maximum likelihood version of {\it ar} of R selects the first 12 lags and has a residual sum of squares of 1583885. 

The American Business Cycle \cite{ABC86} data consists of quarterly data from 1919-1941 of the variables GNP72, CPRATE, CORPYIELD, M1, M2, BASE, CSTOCK, WRICE67, PRODUR72, NONRES72, IRES72, DBUSI72, CDUR72, CNDUR72, XPT72, MPT72, GOVPUR72, NCSPDE72, NCSBS72, NCSCON72, CCSPDE72 and CCSBS72 available from
\begin{verbatim}
http://data.nber.org/data/abc/
\end{verbatim}
In Table~\ref{tab:abcq_lag} the dependent variable is GNP72 (Gross National Product in 1972 dollar terms), the covariates are the lags 1:16 of all the 22 variables. These are ordered as follows. The first 16 covariates are the 16 lags of the first variable GNP. The second sixteen are the 16 lags of the second variable CPRATE and so on. Thus covariate 180 is the fourth lag of the 11th variable  IRES72, $180=11*16+4$.

 In contrast to all other data sets {\it desparsified lasso} and {\it ridge.proj} select a large number of covariates, {\it EMVS} selects 223 which together with the intercept give a perfect fit. Of the model based procedure {\it SIS} is the only one which gives a reasonable selection. All the five selected covariates have Gaussian P-values less then 0.05, but two have values exceeding 0.01. If we run {\it f1st} with $p0=0.05$ it returns six covariates with a sum of squared residuals of 16870.

The analysis can be repeated for the remaining 21 variables giving a vector autoregressive scheme. The results for {\it f1st} are given in Table~\ref{tab:vec_lag} with $p0=0.01$ replaced by $p0=0.01/22$.  

\begin{table}[hb]
\begin{center}
\begin{tabular}{rccc}
\multicolumn{4}{c}{Sunspot data (3253,3252)}\\
\hline\\
method&$k,fp$&ssq.resid&time\\
\hline\\
{\it f1st}&54,0&2023014&1(2.41)\\
{\it lasso}&787,649&632108&11\\
{\it scalreg}&62&2095413&1285\\
{\it SIS}&*&*&*\\
{\it stability}&34,3&2854149&15\\
{\it multi.split}&57&2027120&400\\
{\it BoomSpikeSlab}&52&4630928&24\\
{\it EMVS}&0&*&20\\
\end{tabular}
\caption{The results for the sunspot data. \label{tab:sunspot}}
\end{center}
\end{table}
\clearpage
\begin{table} [ht]
\begin{center}
\begin{tabular}{rcccc}
\multicolumn{5}{c}{Sunspot, lags (2753,500)}\\
\hline\\
method& $k,fp$&P-values&$\bs{ss}$&time\\
\hline\\                     
{\it f1st}&7,0&yes& 1507616&1 (0.436)\\
{\it f3st,m=1}&7,0&yes&1507616&7\\
{\it lasso}&11,4&no&1522261&35\\
{\it scalreg}&451,392&no&1255481&32\\
{\it SIS}&314,286&no&1263136&825\\
{\it desparsified lasso}&7,0&yes&1557744&14853\\
{\it stability}&8,1&no&1545191&30\\
{\it ridge.proj}&5,0&yes&1579760&82\\
{\it multi.split}&7,0&yes&1546039&3403\\
{\it EMVS}&500,0&no&1254916&6\\
{\it BoomSpikeSlab}&(7,0)&no&1956351&20\\
\end{tabular}
\caption{The results for the sunspot data with lags 1:500. \label{tab:sunspot_lag}}
\end{center}
\end{table}

\begin{table} [ht]
\begin{center}
\begin{tabular}{rcccc}
\multicolumn{5}{c}{American Business Cycle (224,352)}\\
\hline\\
method& $k,fp$&P-values&$\bs{ss}$&time\\
\hline\\                     
{\it f1st}&3,0&yes& 18765&1 (0.039)\\
{\it f3st,m=1}&6,0&yes&18405&3\\
{\it lasso}&4,2&no&24980&3\\
{\it scalreg}&83,69&no&4960&19\\
{\it SIS}&5,0&no&17854&16\\
{\it desparsified lasso}&186,185&yes&1880&1000\\
{\it stability}&2,0&no&25460&12\\
{\it ridge.proj}&102,95&yes&8130&30\\
{\it multi.split}&2,0&yes&25460&200\\
{\it EMVS}&223, NaN&no&0&2.8\\
{\it BoomSpikeSlab}&(4,0)&no&48750&65\\
\end{tabular}
\caption{The results for the American Business Cycle data with lags 1:16. \label{tab:abcq_lag}}
\end{center}
\end{table}
\clearpage
\begin{table} [ht]
\begin{center}
\begin{tabular}{ccccc}
\multicolumn{5}{c}{American Business Cycle}\\
\multicolumn{5}{c}{Vector autoregression  (224,352)}\\
variable&\multicolumn{4}{c}{selected covariates}\\
    1 &   1&   18 & 180&*\\
    2&   17&  161 & *&*\\
    3 & 33&   17&*&*\\
    4  & 49& *&*&*\\
    5 &  65&*&*&*\\
    6 &  81 &*&*&*\\
    7&   97 &  98&*&*\\
    8 & 113&  117&   69&*\\
    9 & 129&  193&*&*\\
   10 & 145&*&*&*\\
   11 & 161 & 162&*&*\\
   12 & 177&*&*&*\\
   13 & 193 &  17 &  21&*\\
   14 & 209 &*&*&*\\
   15 & 225&  193 &  12&*\\
   16 & 241&*&*&*\\
   17 & 257&  258&*&*\\
   18 & 273&  129&*&*\\
   19&  289 & 145 & 161&*\\
   20 & 305&  193 &  17&   26\\
   21 & 321 & 323 & 322&*\\
    22 & 337&  326&*&*\\
\end{tabular}
\caption{American Business Cycle, vector autoregressive scheme with lags 1:16. \label{tab:vec_lag}}
\end{center}
\end{table}

 \subsection{Constructing models}
The residuals of the riboflavin data look more or less normal so the data can be modelled with the selected covariates plus Gaussian noise.

 Data such as the leukemia data are usually analysed using the logit model. This can be done by taking the covariates selected for the linear approximations using the Gaussian step-wise procedure and running a logistic regression using these covariates.  If this is done for the 142 linear approximations returned by {\it f2st} 32 of them give a perfect fit.

The sunspot data can be modelled as follow. Let $s(t)$ be the number of sunspots at time $t$ and denote by $f_1$ the function returned by the step-wise procedure. The residuals are defined as $r_1=s-\log(1+\exp(f_1))$ where the transformation is made to force the function to be non-negative. They are reasonably described by an autoregressive process of order 1 with coefficient 0.4. We write
\[d(t)=s(t)-0.4s(t-1), t=2,\ldots,3253\]
 which removes the autocorrelation. We now apply the step-wise procedure to $d$ with the covariates $tr_j(t),t=2,\ldots,3253, j=1,\ldots,3252$ and denote the resulting function by $f_2$. This eliminates the high frequency trigonometric functions caused by the autocorrelations. We define a new approximating function by
\[f_3(t)=\sum_{j=0}^60.4^jf_2(t-j)\]
with residuals $r_3=s-\log(1+\exp(f_3))$. The size of the residuals depends on $\log(1+\exp(f_3))$ and a simple linear regression shows that 
\[z=r_3/(1.28(9.4+0.125\log(1+\exp(f_3)))\]
can be well approximated by an autoregressive Gaussian process of order one with parameter 0.5. This leads to the model
\[S(t)=f_3(t)+1.28(9.4+0.125\log(1+\exp(f_3)))Z(t)\]
where $Z(t)\sim N(0,1)$ is a Gaussian autoregressive process of order with parameter 0.5.

\subsection{Graphs}
There are several methods for calculating dependency graphs. The ones we compare here are the Gaussian covariate graph function {\it fgr1st}, {\it thav.glasso} (\cite{LASZ21}), {\it huge}  (\cite{JFLRLWLZ21}) and {\it glasso} (\cite{FRHATI19}).
 
We consider firstly a random graph generated as in \cite{MEIBUE06} but with the correction given in \cite{DAVDUEM22}. The undirected graph of the Gaussian procedure was used. The procedures {\it thav.glasso} and {\it glasso} have regularization parameters {\it C} and {\it rho} respectively with no indication as to how to choose them. They were chosen to equalize approximately the numbers of false positives and false negatives. The values used were {\it C=0.4} and {\it rho=0.236}. Table~\ref{tab:rnd_graph} gives the results. The second column gives the number of edges, the third the number of false positives $fp$, the fourth the number of false negatives $fn$ and the fifth  the time in seconds. 
\begin{table}
\begin{center}
\begin{tabular}{rcccc}
\multicolumn{5}{c}{Random graph (1000,1000)}\\
\hline\\
method&no. edges&$fp$&$fn$&time\\
\hline\\
{\it fgr1st} &1821&1&3&25\\
{\it thav.glasso} &1776&218&265&90\\
{\it huge}&1830&26&19&30\\
{\it glasso}&1840&293&276&16\\
\end{tabular}
\caption{The results for one simulation of the random graph. \label{tab:rnd_graph}}
\end{center}
\end{table}

The results for the covariates of the riboflavin data were as follows. The procedures {\it thav.glasso} and {\it glasso} were killed after one hour with no results. There may be values of the regularization parameters which work but the user is given no hint. The {\it huge} procedure took 35 seconds but returned zero edges. The Gaussian covariate procedure with the default values took 16 seconds and yielded a directed graph with 4491 edges and an undirected graph with 3882 edges.

\subsection{Summary}
The above demonstrates the overwhelming superiority of the Gaussian covariate procedure. This is exemplified by the two simulations Tables~\ref{tab:riboflavin} and \ref{tab:leukemia} where there is a true model which all but {\it BoomSpikeSlab} fail to find, Table~\ref{tab:riboflavin} where the best in terms of sum of squared residuals {\it scalreg} would be number 88 in the complete list of Table~\ref{tab:f3st}, and the Boston Housing data Table~\ref{tab:boston} where only five of the ten model based procedures got so far, three of these resulted in errors and the remaining two remaining failed both with respect to the selection of covariates and the times required. This may be compared with Table~\ref{tab:Boston_int} in the Appendix.

\section{Reasons for the failure of model based procedures}
\subsection{Truth, P-values and confidence intervals} \label{sec:true_etc}

The decision as to whether a given covariate $\x_j$ lies in  ${\mathcal S}_0$ is made by formulating and calculating a P-value for the hypothesis $H_j: \beta_j=0$.  Here is an example taken from \cite{LOCK17}. On page 5 the covariate $\x_{1278}$ is considered and regressing the dependent riboflavin variable on $\x_{1278}$ results in the $P_F$-value 9.06e-10. This is the smallest value amongst all covariates. It is not however included amongst the selected covariates of Table~\ref{tab:gaucov_results}. The reason is that when regressed together with the four other covariates in Table~\ref{tab:gaucov_results} its $P_G$-values is 0.2474591 which exceeds the threshold of $p0=0.01$. The covariate $\x_{73}$  exhibits the opposite behaviour. Its $P_F$-value when regressed alone on the dependent variable is 2.051118e-02 but its $P_F$-value when regressed with the covariates $\x_{400}, \x_{2564}$ and $\x_{2034}$ is 9.968584e-13.

Which $P_F$-values are correct? Under the model the only correct $P_F$-values are those for the subset ${\mathcal S}_0$, all others are wrong. All P-value correction methods such as Bonferroni assume that the P-values are correct. It is therefore not surprising that the corrected P-values are of no help. But even if they are correct the corrected P-values may still be of no help. It we take the first line of Table~\ref{tab:f3st} the smallest $P_F$-value is 1.83e-23 for the covariate $\x_{4004}$. There are   3.766288e+21 subsets of size seven leading to a corrected $P_F$-value of 0.068 (see page 19 of \cite{LOCK17}) which is not very convincing. The corresponding $P_G$-value is 7.42e-20 which is correct as it stands.

All the above holds  mutatis mutandis for confidence intervals. A confidence interval for a parameter $\beta_i$ is an interval which contains the true value, that is,  the value of $\beta_i$ under which the dependent variable $\Y$ was generated, with a specified probability.  For this it is necessary to know the true subset ${\mathcal S}_0$.

All P-values calculated for data generated under the model (\ref{equ:model}) have the same source of randomness, namely ${\bs \varepsilon}$. Thus not only are the resulting P-values in general false, they are also dependent further undermining their usefulness data analysis.

The conclusion is that concepts such as F P-values and confidence intervals derived from modelling with all covariates cannot be transferred to the problem of determining the correct model even when such a model exists.

\subsection{Data splitting}
The above mentioned problems with P-values and confidence intervals are well known. One method of trying to overcome them is data splitting. We follow here  \cite{DEBUEMEME15} but see also \cite{WASSROED09}. Split the sample into two $I_1$ and $I_2$  at random with $I_1$ of size $n/2$. Based on $I_1$ select a subset $\tilde{{\mathcal S}}_1$ of the covariates and calculate the P-values of the covariates in $\tilde{{\mathcal S}}_1$ using $I_2$. For covariates not in $\tilde{{\mathcal S}}_1$ the P-value is set to 1. This is repeated about 100 times and the P-value for each covariate is calculated from the 100 P-values, for example twice the median. For the procedure to work ${\mathcal S}_0\subset \tilde{{\mathcal S}}_1$ must hold with high probability. There are various suggestion as to how to choose the subset $\tilde{{\mathcal S}}_1$ (see \cite{WASSROED09}) but in the end it does not work. Four applications of {\it multi.split} to the riboflavin data resulted in the covariates  $(\x_{4003})$, $(\x_{1516},\x_{4003})$, $ (\x_{1278},\x_{4003})$ and $(\x_{4003})$ whose corrected P-values varied from 0.019 to 0.984. As a comparison the $P_G$-value of $\x_{4003}$ alone is 7.88e-05.

\subsection{Lasso}
Lasso is one of most used selection procedures and is an integral part of others, for example {\it stability}. Nevertheless lasso does not work and one reason for this may well be that it shrinks the coefficients, one interpretation of which  is that it penalizes the best. We consider the riboflavin data. The first nine covariates on the lasso path  are in order (see \cite{LOCK17} page 17)
\[\x_{1278},\x_{4003},\x_{1516},\x_{2564},\x_{1588},\x_{624},\x_{1312},\x_{1502},\x_{1639}.\]
The first nine selected by the Gaussian covariate method are, in order
\[\x_{1278},\x_{4003},\x_{2564},\x_{73},\x_{2034},\x_{1131},\x_{1762},\x_{2186},\x_{3459}.\]
If the dependent variable is regressed on both sets of covariates the sum of squared residuals is 9.97 for the lasso covariates and 3.98 for the Gaussian covariates. The Gaussian covariate function {\it f1st} selects the four covariates $\x_{73}, \x_{2034}, \x_{2564}$ and $\x_{4003}$, Table~\ref{tab:gaucov_results}. The sum of squared residuals is 8.448 which is less than that for all nine lasso covariates. The same occurs for the Boston data. With $\lambda=1.2$ lasso returns 13 covariates with sum of squared residuals 11819. The first thirteen covariates using the Gaussian procedure have a sum of squared residuals of 5450. The nine covariates selected by the procedure have a sum of squared residuals of 5589, Table~\ref{tab:Boston_int}. Given this it is perhaps not surprising why lasso fails. The covariates on the lasso path are so to speak, greatly inferior to those on the greedy forward search path. 

There have been attempts to provide valid P-values along the lasso path, see \cite{LOKTAYTIB214} and \cite{TIBTAYLOCTIB16}. There are however problems. On page 19 of  \cite{LOCK17} a significance test for the covariate 1278 of the riboflavin data is considered. The test statistic of \cite{LOKTAYTIB214} for the first covariate  is
\[ T_1=\frac{\lambda_1(\lambda_1-\lambda_2)}{\hat{\sigma}^2}=24 \text{ or }2.55\]
with P-values respectively 0.078 or $3.7\time 10^{-11}$ depending on the choice of $\hat{\sigma}^2$. Lockhart comments `Estimation of $\sigma$ is crucial and hard, I think'. On page 43 we read that one of the $\hat{\sigma}$ values is the standard deviation of the dependent variable. The second is `based on a fit to some 30 predictors'. On page 19 we read `\cite{LOKTAYTIB214} mostly consider a fictitious universe in which $\sigma$ is known'. The same applies to \cite{TIBTAYLOCTIB16}. The problem of estimating $\sigma$ remains.

\subsection{Bayesian procedures} \label{sec:bayes}
The Bayesian paradigm carries over without change to the situation $q>n$. The simplest model is
\begin{equation} \label{equ:bayes1}
\Y=\X{\bs \beta}+\sigma{\bs \varepsilon}
\end{equation}
which must be supplemented with priors for ${\bs \beta}$ and $\sigma$. This has a certain elegance when compared with the collapse of the P-value paradigm but it is not without problems. 

In the riboflavin simulations of Section~\ref{sec:ribo_sim} there was one unique correct model. BoomSpikeSlab was slow but in Table~\ref{tab:ribo_sim} it was the third best procedure in terms of identifying the correct model . In the case of the riboflavin data itself it was the second best of the model based procedures after {\it SIS}  but much worse than {\it f1st} and a factor 2290 slower.  It failed completely for the osteoarthritis (Table~\ref{tab:osteoarthritis}) and for the Boston Housing  (Table~\ref{tab:boston}) data sets. 

All these failures are due to the problems with the prior over a high dimensional space $\R^q$. Firstly there is the algorithmic problem of calculating the posterior.  BoomSpikeSlab uses MCMC making it exceptionally slow. Secondly it is based on the covariance matrix of the covariates which in the case of the osteoarthritis data has 1190842003 entries and a size of about 17Gb. This explains the errors for both Bayesian procedures in Table~\ref{tab:osteoarthritis}. They would be completely overwhelmed by the Boston Housing data whose covariance matrix is of size about  230Gb.

\subsection{The mathematics of covariate selection} \label{sec:math_cov_sel}
The mathematics of model based procedures is not simple, see for example \cite{TIBTAYLOCTIB16}, the Appendix of \cite{DEBUEMEME15} and \cite{BUEH13}. All the P-value based procedures make use of regularization and in particular many make use of lasso. Such procedures requires a specific value of the regularization parameter $\lambda$ and this value depends on the data. All results on choosing $\lambda$ are of asymptotic nature and many require a value for $\sigma$ in (\ref{equ:model}). None of this is any help in choosing $\lambda$ for a specific data set with the consequence that the default method for lasso is cross-validation as it was 26 years ago in \cite{TIB96}, and it fails. 

Contrast this with the Gaussian method which only needs the threshold $ p0$ for the Gaussian P-values. It has an immediate interpretation, the default value $p0=0.01$ was use for all data sets and simulations. By far the most mathematically demanding sections  of \cite{DAVDUEM22} are those on the asymptotic behaviour for a given $p0$ ($\alpha$ in  \cite{DAVDUEM22}). They have nothing to do with the choice of $p0$, they only confirm its interpretation.

\subsection{Memory requirements and algorithmic complexity} \label{sec:mem_alg}
Both {\it EMVS} and {\it BoomSpikeSlab} fail for the osteoarthritis and Boston data sets because of allocation problems alone as shown in the previous Section. For the same reason {\it thav.glasso}, {\it huge} and {\it glasso} also fail for these two data sets as they also require the covariance matrix of the covariates. The memory requirements of {\it f1st} are about $nq$ and those of {\it fgr1st} about $4nq$. 

In \cite{HUB11} we read `Pay attention to computational complexity; keep it below $O(n^{3/2})$, or forget about the algorithm'. To which we can add, if your algorithm requires the covariance matrix of the covariates, forget about it.

The stepwise version of the Gaussian covariate procedure is very fast. If $k$ covariates are chosen it requires about $6nq(k+1)$ multiplications which is a factor of $\sqrt{nq}$ smaller than the complexity given by Huber. This is reflected in the times for the various procedures given above. The fastest of these is {\it EMVS} which is a factor of about 20 slower than Gaussian covariates but only seems to select zero covariates. The algorithmic complexity of the other procedures seems not to be given.

\subsection{Multiple approximations and models} \label{sec:mult_approx}
The model based approach postulates a single model such as (\ref{equ:model}).  The goal is to identify the active subset $ {\mathcal S}_0$. Consequently model based procedures are designed to output just one set of selected covariates. In practice repetitions will result in different selected covariates but this is not intended, it is simply the use of some form of random data splitting. The word true is part of the vocabulary, sometimes in inverted commas `true' to emphasize that the author does not believe it. Articles such as \cite{BENJHOCH95} are completely reliant on it. There is no concept of approximation.

One advantage of the model approach may be thought to be that the problem of covariate choice can be posed as an exact mathematical problem, essentially that of finding the true model (\cite{CAFAJALV2018}).  When thinking  in terms of approximations it makes little sense to talk of the true approximation. In general there will be several adequate approximations and they may well be equally valid. This can be posed as a precise mathematical problem using the concept of Gaussian P-values as follows: find all subsets with the property that no $P_G$-value exceeds $p0=0.01$. If $q$ is not too large the all subset version of the Gaussian covariates procedure solves this problem.

For large $q$ the all subset version fails and there is no general strategy for producing adequate approximations. The greedy forward selection of the Gaussian stepwise procedure is one of the oldest selection procedures. The success of the Gaussian version is due to the stopping rule and the availability of exact and universally valid $P_G$-values. Its disadvantage is that it yields only one approximation. The Gaussian covariate R package offers two functions {\it f2st} and {\it f3st} to overcome this. They are described in the Appendix.

This is not an argument against models, it is an argument against a vocabulary and practice of treating models as true. A model can be explicitly treated as an approximation but this requires a concept of approximation and the definition of some form of adequate approximation. A concept such as confidence regions is no longer defined as a region containing the true parameter values with a given probability, but a region which contains those parameter values which give an adequate approximation (see \cite{DAV14} and also \cite{TUK93B}. 

Models can be obtained from the approximations by modelling the residuals. The residuals of the riboflavin data look more or less like i.i.d. Gaussian with of the model based procedures$\sigma$ set to the standard deviation. For the osteoarthritis data a logit model can be used.  The covariates for the leukemia data can be used for a logit model. The Boston data are more difficult as the residuals exhibit a structure and there are outliers. In \cite{DAVDUEM22} the sunspot data are modelled using sines and cosines with the residuals being an AR(1) Gaussian process with parameter 0.5.

\subsection{Gaussian covariates} \label{sec:gau_cov}
As has been repeatedly stated the Gaussian covariate P-values are exact and valid no matter what the data and what the subset. The memory demands and the algorithmic complexity are linear in the size of the data. These may be necessary conditions for good performance but they are not sufficient. To understand better why it works so well one can investigate conditions under which it can be shown to work. This is done by means of asymptotics in Section~9 of  \cite{DAVDUEM22} which gives  sufficient conditions on $(n,q)$, the size $k^*$ of the active set ${\mathcal S}_{k^*}$  and the sizes of the $\beta_j$ for $\x_j \in  {\mathcal S}_{k^*}$ for the stepwise procedure to work. 

The first step in the stepwise procedures consists of calculating $\max_{1 \le i\le q}\vert \x_i^t\y\vert$ and comparing this with $\max_{1 \le i\le q}\vert \Z_i^t\y\vert$ where the $\Z_i$ are i.i.d. Gaussian covariates. Gaussian covariates can be thought of as the most irrelevant of the universally irrelevant covariates. If $\y=\Z$ is just Gaussian noise and the $\x_i$ have norm 1 it follows from the Gaussian correlation inequality (\cite{ROY14}) that
\[\bs{P} \bigl( \max_{\x_i} (\x_i^t\Z)^2 \ge x \bigr)
	\ \le \ \bs{P} \bigl( \max_i Z_i^2 \ge x \bigr)
\]
with equality when the $\x_i$ are uncorrelated. This is used in the proof of Theorem~3 of \cite{DAVDUEM22}. 

A similar approximate result for fixed $\y$ can be obtained by considering uniform orthogonal rotations of the $\x_i$ (see \cite{DUMDAV20}).  These retain the geometric structure of the covariates. The message is that highly correlated covariates $\x_i$ reduce the effective value of $q$. The extreme example is when the $\x_i$ are all equal in which case the effective value of $q$ is $q=1$. It may be possible to improve the performance of the Gaussian covariate procedure by taking the covariance structure of the $\x_i$ into account but it is not clear how to do this.

\begin{figure}[hb]
  \centering
\title{Correlation structure of covariates of data sets}
\\
\includegraphics[width=.4\textwidth,height=150px]{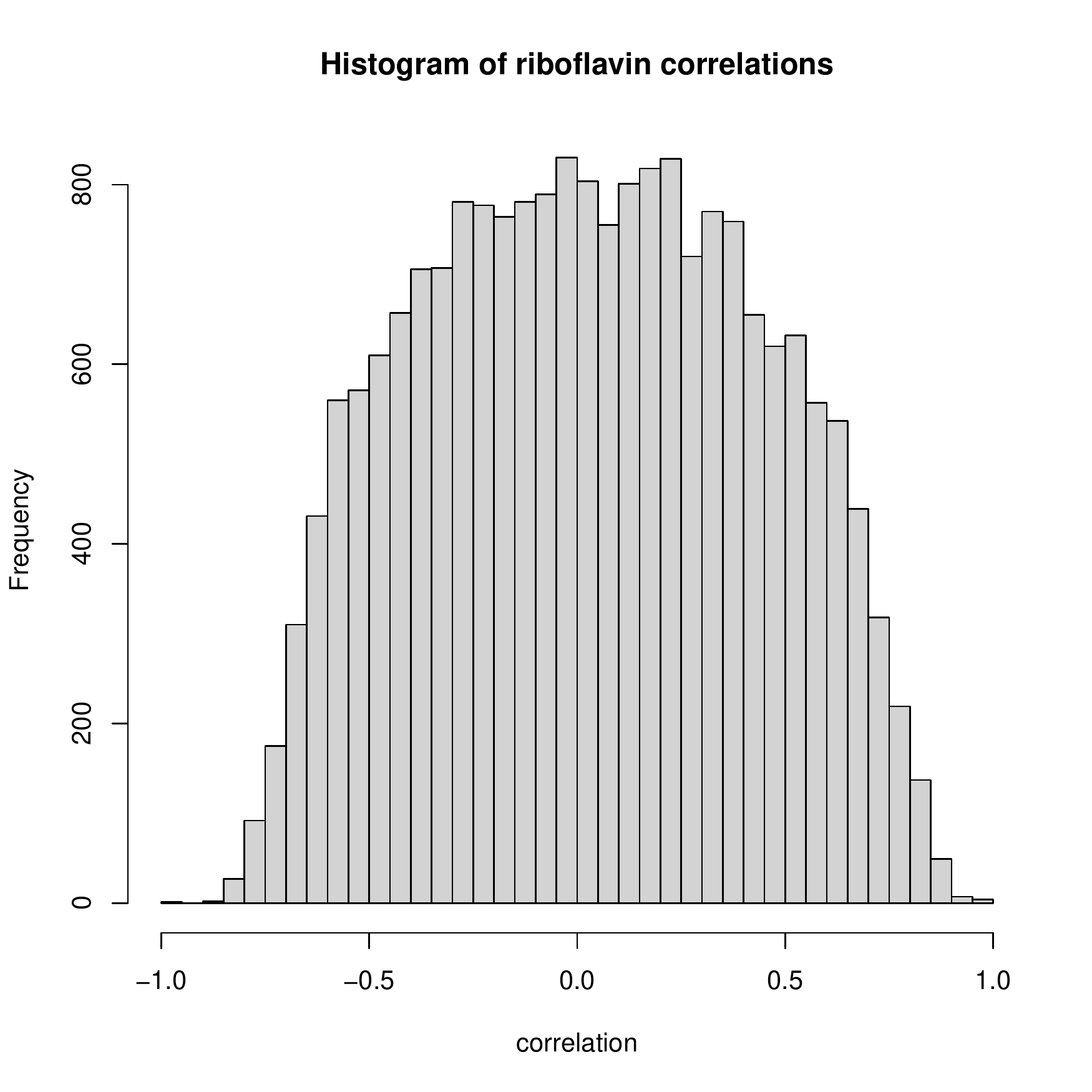}
\includegraphics[width=.4\textwidth,height=150px]{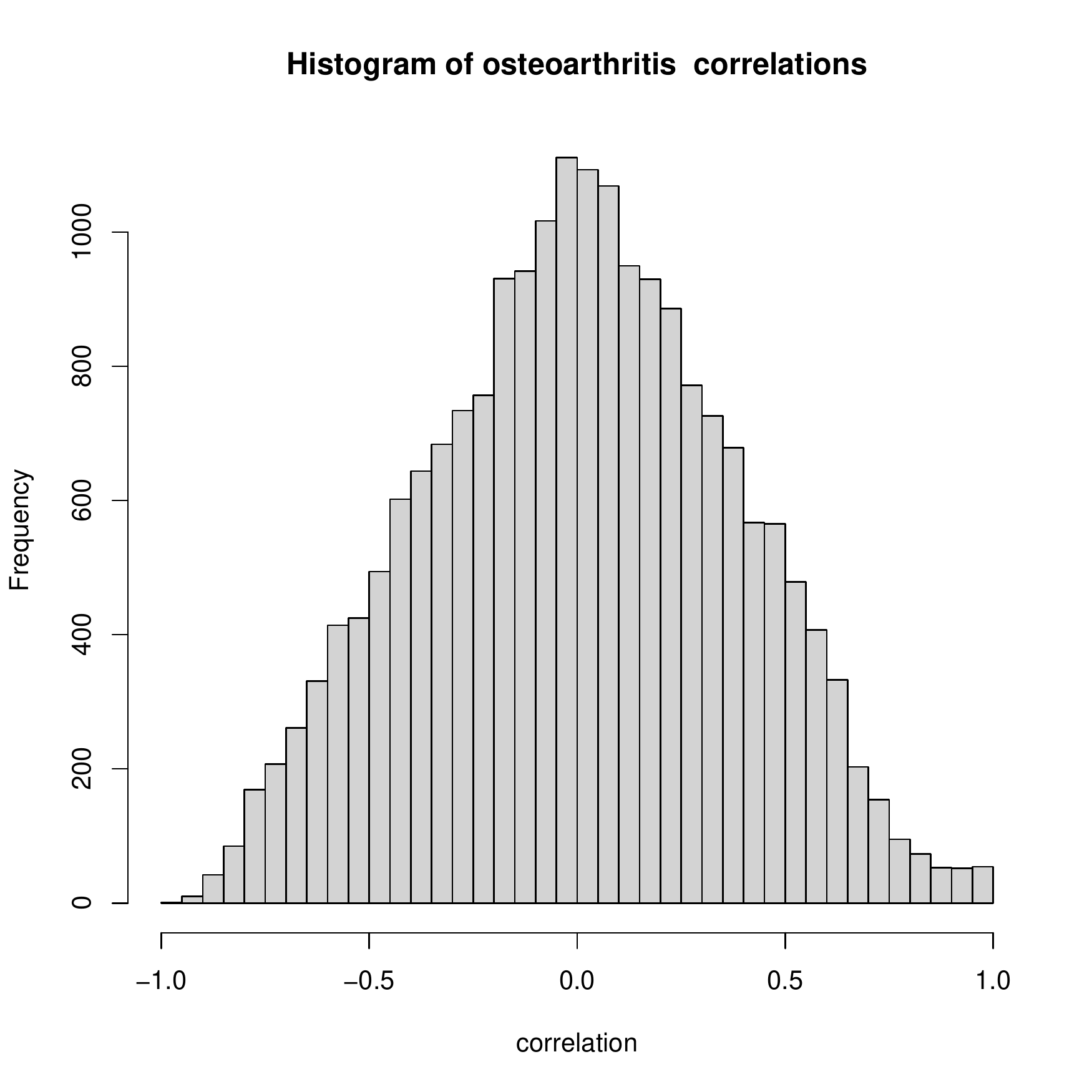}\\

\includegraphics[width=.4\textwidth,height=150px]{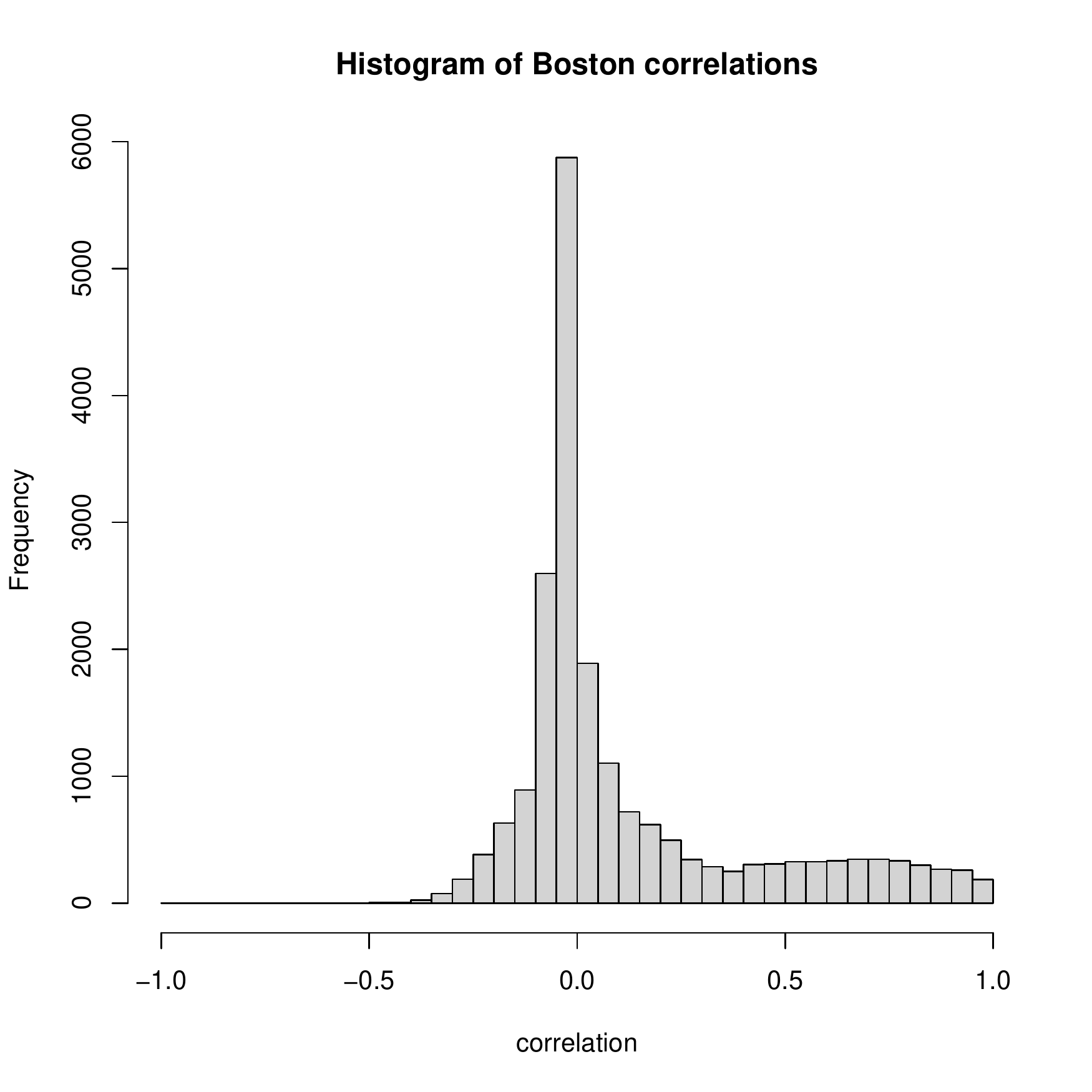}
\includegraphics[width=.4\textwidth,height=150px]{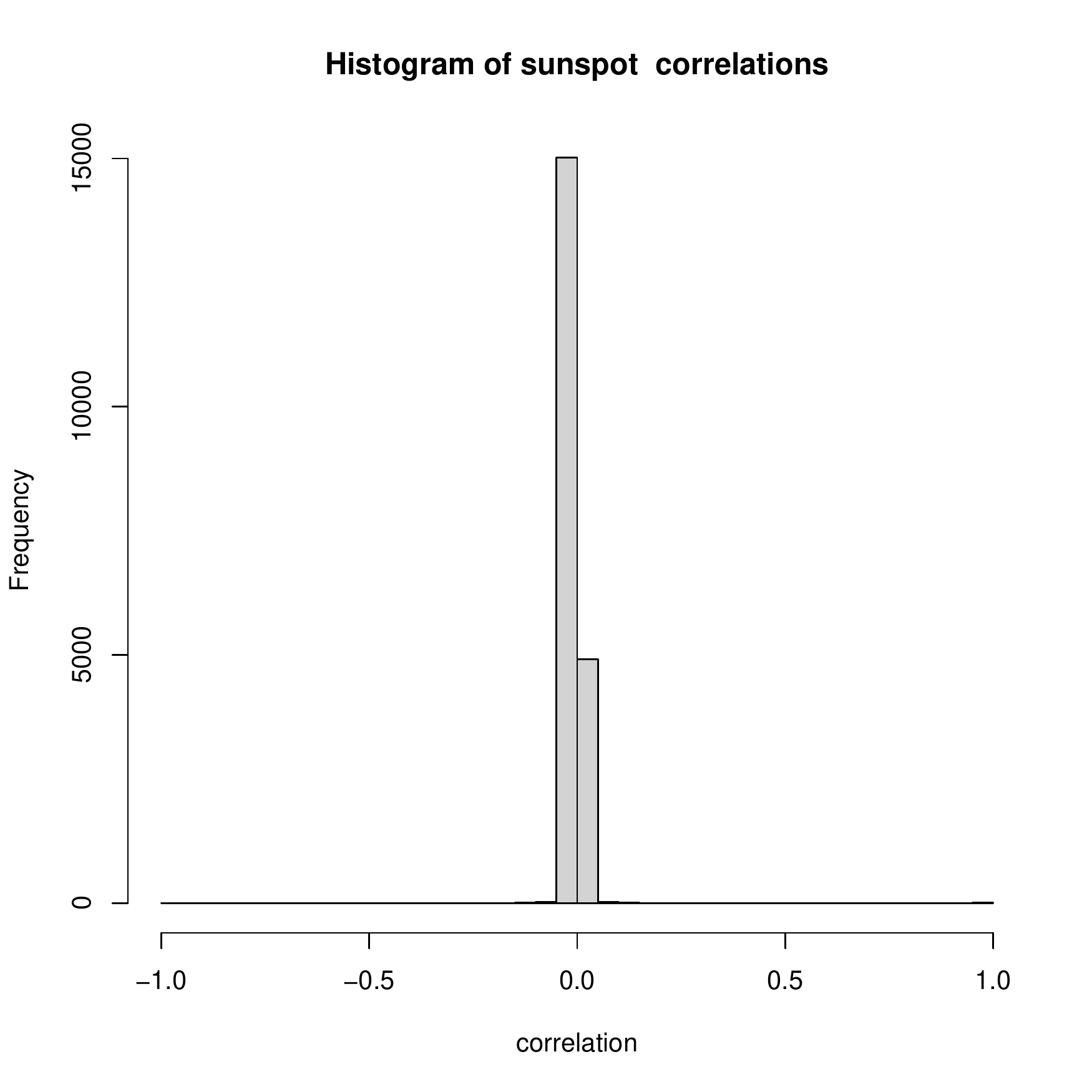}
\caption{From top left clockwise: riboflavin, osteoarthritis, sunspot, Boston housing.
 \label{fig:cor}}
\end{figure}

Figure~\ref{fig:cor} shows the correlation structure of the covariates of the four data sets based on a random choice of 20000 pairs. The Gaussian covariate procedure performs uniformly well over all data sets. The procedures  {\it scalreg, stability, multi.split} and {\it BoomSpikeSlab}  performed best for the sunspot data of Table~\ref{tab:sunspot} in terms of covariates selection. Of these {\it stability} performed the worst as it clearly underfitted. Again the Gaussian covariate procedure performed best in that it selected fewer covariates but nevertheless had the smallest sum of squared residuals. It was also by far the fastest procedure. The other three all failed for the sunspot data: {\it lasso} performed worse for this data set than for the other three, {\it  EMVS} returned zero as for the other data sets with the exception of Table~\ref{tab:sunspot_lag} where it returned all 500, the initial screening in {\it SIS} failed completely presumably because of the near orthogonality of the covariates. 

\subsection{Publications and rejections} \label{sec:publ}
The model based procedures all of which have failed have an impressive record of publication in leading statistical journals. Of those mentioned in this article three were published in the AoS, five in JRSSB, two in JASA, one in Statistical Science  and two in Biometrika. 

The Gaussian procedure, the only one which works,  also has an impressive record  with rejections from AoS (twice), JRSSB as a discussion paper, JASA, EJS and AoAS. Clearly statistical journals prefer mathematically demanding and complicated methods that don't work to simple ones which do.

\subsection{Simplicity}
The Gaussian covariates procedures are based on  (\ref{equ:rss_dist}) which is a special case of Theorem 1 of \cite{DAVDUEM22}. The Gaussian $P_G$ values follow directly from this. All inherit the properties of (\ref{equ:rss_dist}), namely that they are exact and valid no matter what the data, the subset or the covariate. The calculations require only least squares without needing either a QR or Singular Value Decomposition. The only exceptions are the subroutines for the Beta probabilities which are taken from \cite{PRTEVEFL03}. The algorithms  are of linear complexity as are the memory requirements. The result is overwhelmingly superior to all model based procedures. In short,  (\ref{equ:rss_dist})  replaces about 60 years of failed model based research on covariate selection for linear regression. There is no reason why the method could not have been developed 60 years ago, apart from the reluctance of statisticians to consider anything outside the statistical canon (see Section~\ref{sec:publ}).

\section{Appendix} \label{sec:appendix}
This gives a short and simple introduction to the Gaussian covariate procedure of \cite{DAVDUEM22}. In spite of its simplicity it represents a complete break with the current statistical paradigm.

Given a dependent variable $\y$ of size $n$, covariates $\x_i, i=1,\ldots,q$  and subset $\mathcal{S}_k$ of the $\x_i$ of size $k<\min(n,q)$ regress $\y$ on the covariates in   $\mathcal{S}_k$  to give a sum $\rss_k$ of squared residuals. Choose an $\x_i \in \mathcal{S}_k$ and replace it with a Gaussian covariate $Z$ to give a random sum of squared residual $\RSS$. Then
\begin{equation} \label{equ:rss_dist}
RSS/rss_{k,-i} \sim \text{Beta}((n-k)/2,1/2))
\end{equation}
where $\rss_{k,-i}$ is the sum of squared residuals when $\y$ is regressed on $\mathcal{S}_k\setminus \{\x_i\}$.  From this and the identity
\begin{equation} \label{equ:beta_F}
\text{F}_{\nu_1,\nu_2}(x)=\text{Beta}_{\nu_1/2,\nu_2/2}((\nu_1/\nu_2)x/((\nu_1/\nu_2)x+1))
\end{equation}
it follows after some manipulation that the standard P-value of $\Z$ is uniformly distributed over $(0,1)$. Equation (\ref{equ:rss_dist}) is a special case of Theorem~1 of \cite{DAVDUEM22}. It holds whatever the data and whatever the subset.

It follows from (\ref{equ:rss_dist}) and (\ref{equ:beta_F}) that the probability that the Gaussian covariate $\Z$ is better than $\x_i$, that is if $\RSS<\rss_k$, is given by
 \begin{equation} \label{equ:gau_pval_0}
 \P(\RSS<\rss_k)=\text{Beta}_{(n-k)/2,1/2}(\rss_k/\rss_{k,-i})=P_F(\x_i)
\end{equation}
where $P_F(\x_i)$ is the standard F P-value based on the subset $\mathcal{S}_k$.

Any subset $\mathcal{S}_k, k<n$ of the covariates specifies a selection of covariates and defines  a linear approximation to $\y$ obtained by regressing $\y$ on the covariates in $\mathcal{S}_k$. When selecting covariates the problem is to avoid including ones which are irrelevant, defined here  as being no better than Gaussian noise.  This will be done by defining a Gaussian P-value as follows. Given a subset $\mathcal{S}_k$ size $k$ and a covariate $\x_i \in \mathcal{S}_k$ we replace $\x_i$ and the remaining $q-k$ covariates by $q-k+1$ by i.i.d. Gaussian covariates  $\Z_j, j=1,\ldots,q-k+1$ . For each $\Z_j$ we regress $\y$ on $\mathcal{S}_k\setminus \{\x_i\}\cup\{\Z_j\}$ to give a sum of squared residuals  $\RSS_{k-1,+j}$. The Gaussian P-values of $\x_i$ should measure the degree to which $\x_i$ is better than the best of the Gaussian covariates $\Z_j$. This is accomplished by defining the the P-value of each $\x_i$ by
\begin{eqnarray}
P_G(\x_i)&=&\P(\min_j\RSS_j<\rss_k)=\text{Beta}_{1,q-k+1}(\text{Beta}_{(n-k)/2,1/2}(\rss_k/\rss_{k,-i}))  \label{equ:gau_pval_1}             \\
&=&1-(1-P_F(\x_i))^{q-k+1}.\nonumber
\end{eqnarray}
The second equality follows form the independence of the $\RSS_{k-1,+j}$ and (\ref{equ:gau_pval_0}).

The all subset version of Gaussian covariate selection  retains only those subsets for which each covariate $\x_i$ in the subset has a $P_G$ value (\ref{equ:gau_pval_1}) $P_G(\x_i) <p0$. The procedure in {\it gausscov} further eliminates all retained subsets which are subsets of some other retained subset. Those left are ordered according to the sum of squared residuals. These subsets are maximal in the sense that it is not possible to include another covariate and still retain the upper bound $p0$. Clearly the all subset procedure is only possible if $q$ is not too large, say $q\le 25$.

The stepwise selection procedure based on Gaussian covariates makes use of a slightly different $P_G$-value. If a subset $\mathcal{S}_k$ of size $k$ with sum of squared residuals $\rss_k$ has already been selected the next candidate for selection is $\x_b$ the best of the remaining $q-k$ covariates. Its $P_G$-value is given by
\begin{eqnarray}
P_G(\x_b)&=&\P(\min_{\ell}\RSS_{k,+\ell}<\rss_{k,+b})=\text{Beta}_{1,q-k}(\text{Beta}_{(n-k-1)/2,1/2}(\rss_b/\rss_k)) \nonumber\\
&=&1-(1-P_F(\x_b))^{q-k} \label{equ:gau_pval_2}
\end{eqnarray}
where $\rss_{k,+b}$ is the sum  of squared residuals when $\y$ is regressed om $\mathcal{S}_k\cup\{\x_b\}$ and $\RSS_{k,+\ell}$ is the sum  of squared residuals when $\y$ is regressed om $\mathcal{S}_k\cup\{\Z_{\ell}\}$ 

The stepwise selection procedure uses the $P_G$-values (\ref{equ:gau_pval_2}). If a subset $\mathcal{S}_k$ has already been selected the best of the remaining covariates is considered and if its $P_G$-value is less than $p0$ it is selected and the procedure continues. Otherwise the procedure terminates.  If the size of the selected subset does not exceed an upper bound specified by the user all subsets of the selected subset  are considered as above and the best subset in the sense of least squares is returned. The default maximum size for all subsets is 20. If this is exceeded the selected subset is returned with the $P_G$-values.

If the  intercept is specifically included (the default case) it is denoted $\x_0$ in the output,it is always the last covariate and its $P_G$-value is the standard $P_F$-value.

It can happen with the stepwise procedure that it fails at the first step as the first $P_G$-value exceeds $p0$. This is the case for the dental data of \cite{SEHTUK01}. This can be avoided to some extent by specifying a minimum size $kmn$ for the initial selected subset. We give some examples below.

Two further functions are {\it f2st} and {\it f3st}. The first {\it f2st} makes an initial selection with {\it f1st}, eliminates the selected covariates  and applies {\it f1st} to those remaining. This is continued until {\it f1st}  selects no more covariates.  If it is applied to the riboflavin data with {\it kmn=10} it results in 44 approximations involving 132 covariates. The time required is about one second.

The second {\it f3st} again the initial selection is made with {\it f1st} but now each of the selected covariates is eliminated in turn whilst maintaining the others. {\it f1st} is applied again and this it iterated $m$ times where $m$ is specifies by the user.  The initial selection for the riboflavin data consists of the covariates $\x_{73}, \x_{2034}, \x_{2564},\x_{4003}$  with a sum of squared residuals 8.45, Table~\ref{tab:gaucov_results}.  Eliminating $\x_{73}$ results in $\x_{4003},\x_{2564},\x_{143},\x_{2034}$ with sum of squared residuals 9.41,  eliminating $\x_{4003}$ results in $\x_{73}, \x_{1131}, \x_{1278}, \x_{2140}, \x_{2564},\x_{4006}$  with a sum of the squared residuals 6.21.

We illustrate the stepwise procedure {\it f1st} using the Boston Housing data where the covariates are all interactions of order eight and less. The dimensions are $(n,q)=(506,176358)$. The results are given in Table~\ref{tab:Boston_int} for various values of $kmn$.
\begin{table}[h]
\begin{center}
{\footnotesize
\begin{tabular}{rccccccccccc}
\multicolumn{12}{c}{Boston Housing data (506,176358)}\\
\hline
$kmn$&\multicolumn{9}{c}{covariates}&$\rss$&time\\
0&7222&170037&164793&106761&82567&81454&&&&6566&2.94\\
10&7222&164793&106761&81454&29154&170455&67241&&&6130&4.59\\
15&7222&164793&106761&81454&29154&170455&67241&92&23339&5589&6.50\\
17&7222&164793&106761&81454&29154&170455&92&23339&170246&4711&10.1\\
&5862&5927&169947&&&&&&&\\
\end{tabular}
}
\caption{The stepwise procedure with $p0=0.01$ and for various values of the size $kmn$ of the initial selection. The time is given in seconds. \label{tab:Boston_int}}
\end{center}
\end{table}

\section{Acknowledgements}
Thanks to John Hodgson for pertinent comments on earlier versions. He is not responsible for any nonsense still standing.

{}

\begin{thebibliography}{}
\bibitem[Alizadeh et~al., 2000]{ALI00}
Alizadeh, A., Eisen, M., Davis, R., Ma, C., Lossos, I., Rosenwald, A.,
  Boldrick, J., Sabet, H., Tran, T., Yu, X., Powell, J., Yang, L., Marti, G.,
  Moore, T., Hudson, J.~J., Lu, L., Lewis, D., Tibshirani, R., Sherlock, G.,
  Chan, W., Greiner, T., Weisenburger, D., Armitage, J., Warnke, R., Levy, R.,
  Wilson, W., Grever, M., Byrd, J., Botstein, D., Brown, P., and Staudt, L.
  (2000).
\newblock Distinct types of diffuse large b-cell lymphoma identified by gene
  expression profiling.
\newblock {\em Nature}, 403:503--511.

\bibitem[Barber et~al., 2020]{BCJPS20}
Barber, R., Candes, E., Janson, L., Patterson, E., and Sesia, M. (2020).
\newblock The knockoff filter for controlled variable selection.
\newblock https://CRAN.R-project.org/package=knockoff.

\bibitem[Benjamini and Hochberg, 1995]{BENJHOCH95}
Benjamini, Y. and Hochberg, Y. (1995).
\newblock Controlling the false discovery rate: A practical and powerful
  approach to multiple testing.
\newblock {\em Journal of the Royal Statistical Society: Series B},
  57:289--300.

\bibitem[B\"uhlmann, 2013]{BUEH13}
B\"uhlmann, P. (2013).
\newblock Statistical inference in high dimensional models.
\newblock {\em Bernoulli}, 19:1212--1242.

\bibitem[B\"uhlmann et~al., 2014]{BUEKALMEI14}
B\"uhlmann, P., Kalisch, M., and Meier, L. (2014).
\newblock High-dimensional statistics with a view toward applications in
  biology.
\newblock {\em Annual Review of Statistics and Its Applications},
  1(1):255--278.

\bibitem[Cand\`{e}s et~al., 2018]{CAFAJALV2018}
Cand\`{e}s, E., Fan, Y., Janson, L., and Lv, J. (2018).
\newblock Panning for gold: `model{-}{X}' knockoffs for high dimensional
  controlled variable selection.
\newblock {\em JRSSB}, 80(3):551--577.

\bibitem[Cox and Battey, 2017]{COXBATT17}
Cox, D.~R. and Battey, H.~S. (2017).
\newblock Large numbers of explanatory variables, a semi-descriptive analysis.
\newblock {\em Proc. Natl. Acad. Sci. USA}, 114(32):8592~8595.

\bibitem[Davies, 2014]{DAV14}
Davies, L. (2014).
\newblock {\em Data Analysis and Approximate Models}.
\newblock Monographs on Statistics and Applied Probability 133. CRC Press.

\bibitem[Davies and D\"umbgen, 2022]{DAVDUEM22}
Davies, L. and D\"umbgen, L. (2022).
\newblock Covariate selection based on an model-free approach to linear
  regression with exact probabilities.
\newblock arxiv.org/2202.01553.

\bibitem[Dettling and B\"uhlmann, 2002]{DETBUH02}
Dettling, M. and B\"uhlmann, P. (2002).
\newblock Supervised clustering of genes.
\newblock {\em Genome Biology}, 3(2):1--15.

\bibitem[Dezeure et~al., 2015]{DEBUEMEME15}
Dezeure, R., B\"uhlmann, P., Meier, L., and Meinshausen, N. (2015).
\newblock High-dimensional inference: confidence intervals, p-values and
  {R}-software hdi.
\newblock {\em Statistical Science}, 30(4):533--558.

\bibitem[D\"umbgen and Davies, 2020]{DUMDAV20}
D\"umbgen, L. and Davies, L. (2020).
\newblock Connecting model-based and model-free approaches to linear least
  squares regression.
\newblock arxiv.org/abs/1807.09633.

\bibitem[Fan and Lv, 2008]{FANLV08}
Fan, J. and Lv, J. (2008).
\newblock Sure independence screening for ultrahigh dimensional feature space.
\newblock {\em Journal of the Royal Statistical Society: Series B},
  70(5):849--911.

\bibitem[Feng et~al., 2020]{FFSS20}
Feng, Y., Fan, J., Saldana, D., and Samworth, R. (2020).
\newblock Sis: Sure independence screening.
\newblock https://CRAN.R-project.org/package=SIS.

\bibitem[Friedman et~al., 2008]{FHT08}
Friedman, J., Hastie, T., and Tibshirani, R. (2008).
\newblock Sparse inverse covariance estimation withthe graphical lasso.
\newblock {\em Biostatistics}, 9(3):432--441.

\bibitem[Friedman et~al., 2019]{FRHATI19}
Friedman, J., Hastie, T., and Tibshirani, R. (2019).
\newblock Graphical lasso: Estimation of gaussian graphical models.
\newblock https://CRAN.R-project.org/package=glasso.

\bibitem[Friedman et~al., 2021]{FRHATINATASNQI21}
Friedman, J., Hastie, T., Tibshirani, R., Narasimhan, B., Tay, K., Simon, N.,
  and Qian, J. (2021).
\newblock Lasso and elastic-net regularized generalized linear models.
\newblock https://CRAN.R-project.org/package=glmnet.

\bibitem[Golub et~al., 1999]{GOLETAL99}
Golub, T., Slonim, D., P., T., Huard, C., Gaasenbeek, M., Mesirov, J., Coller,
  H., Loh, M., Downing, J., Caligiuri, M., Bloomfield, C., and Lander, E.
  (1999).
\newblock Molecular classification of cancer: class discovery and class
  prediction by gene expression monitoring.
\newblock {\em Science}, 286(15):531--537.

\bibitem[Gordon, 1986]{ABC86}
Gordon, R., editor (1986).
\newblock {\em The American Business Cycle: Continuity and Change}, volume~25
  of {\em National Bureau of Economic Research Studies in Business Cycles}.
\newblock Univerisity of Chicago Press.

\bibitem[Harrison and Rubinfeld, 1978]{HARRUB78}
Harrison, D. and Rubinfeld, D. (1978).
\newblock Hedonic prices and the demand for clean air.
\newblock {\em J. Environ. Economics and Management}, 5:81--102.

\bibitem[Huber, 2011]{HUB11}
Huber, P.~J. (2011).
\newblock {\em Data Analysis}.
\newblock Wiley, New Jersey.

\bibitem[Jiang et~al., 2021]{JFLRLWLZ21}
Jiang, H., Fei, X., Liu, H., Roeder, K., Lafferty, J., Wasserman, L., Li, X.,
  and Zhao, T. (2021).
\newblock huge: High-dimensional undirected graph estimation.
\newblock https://CRAN.R-project.org/package=huge.

\bibitem[Laszkiewicz, 2021]{LASZ21}
Laszkiewicz, M. (2021).
\newblock Thresholded adaptive validation: Tuning the graphical lasso for graph
  recovery.
\newblock https://github.com/MikeLasz/thav.glasso.

\bibitem[Laszkiewicz et~al., 2021]{LFL21}
Laszkiewicz, M., Fischer, A., and Lederer, J. (2021).
\newblock Thresholded adaptive validation:tuning the graphical lasso for graph
  recovery.
\newblock https://arxiv.org/pdf/2005.00466.pdf.

\bibitem[Lockhart, 2017]{LOCK17}
Lockhart, R. (2017).
\newblock Inference in high-dimensional linear models course notes.
\newblock httpDimensionals://www.sfu.ca/~lockhart/richard/Cambridge/Notes.pdf.

\bibitem[Lockhart et~al., 2014]{LOKTAYTIB214}
Lockhart, R., Taylor, J., Tibshirani, R.~J., and Tibshirani, R. (2014).
\newblock A significance test for the lasso.
\newblock {\em Ann. Statist.}, 42(2):413--468.

\bibitem[Meier et~al., 2021]{MDMMB21}
Meier, L., Dezeure, R., Meinshausen, N., Maechler, M., and Buehlmann, P.
  (2021).
\newblock hdi: High-dimensional inference.
\newblock https://CRAN.R-project.org/package=hdi.

\bibitem[Meinshausen and B\"uhlmann, 2006]{MEIBUE06}
Meinshausen, N. and B\"uhlmann, P. (2006).
\newblock High-dimensional graphs and variable selection with the lasso.
\newblock {\em Annals of Statistics}, 34(3):1436--1462.

\bibitem[Meinshausen and B\"uhlmann, 2010]{MEIBUE10}
Meinshausen, N. and B\"uhlmann, P. (2010).
\newblock Stability selection.
\newblock {\em Journal of the Royal Statistical Society: Series B},
  72:1436--146.

\bibitem[Press et~al., 2003]{PRTEVEFL03}
Press, W.~H., Teukolsky, S.~A., Vetterling, W.~T., and Flannery, B.~P. (2003).
\newblock {\em Numerical Recipes in Fortran 77: The Art of Scientific
  Computing}, volume~1.
\newblock Cambridge University Press, second edition.

\bibitem[Rockova, 2021]{ROCMOR21}
Rockova, V.and~Moran, G. (2021).
\newblock Emvs: The expectation-maximization approach to bayesian variable
  selection.
\newblock https://CRAN.R-project.org/package=EMVS.

\bibitem[Rockova and George, 2014]{ROCGEO14}
Rockova, V. and George, E. (2014).
\newblock The em approch to {B}ayesian variable selection.
\newblock {\em Journal of the American Statistical Association}, 109(506).

\bibitem[Royen, 2014]{ROY14}
Royen, T. (2014).
\newblock A simple proof of the {G}aussian correlation conjecture extended to
  some multivariate gamma distributions.
\newblock {\em Far East Journal of Theoretical Statistics}, 48(2):139--145.

\bibitem[Scott, 2021]{SCO21}
Scott, S. (2021).
\newblock Boomspikeslab: {MCMC} for spike and slab regression.
\newblock https://cran.r-project.org/web/packages/BoomSpikeSlab/index.html.

\bibitem[Seheult and Tukey, 2001]{SEHTUK01}
Seheult, A.~H. and Tukey, J.~W. (2001).
\newblock Towards robust analysis of variance.
\newblock In Saleh, A.~K.~M.~E., editor, {\em Data Analysis from Statistical
  Foundations: A Festschrift in Honor of the 75th Birthday of D.A.S. Fraser},
  pages 217--244. Nova Science, New York.

\bibitem[SILSO, 2020]{SIDB}
SILSO (2020).
\newblock The international sunspot number.
\newblock International Sunspot Number Monthly Bulletin and online catalogue.
\newblock Royal Observatory of Belgium, avenue Circulaire 3, 1180 Brussels,
  Belgium.

\bibitem[Sun, 2019]{SUN19}
Sun, T. (2019).
\newblock Scaled sparse linear regression.
\newblock https://CRAN.R-project.org/package=scalreg.

\bibitem[Sun and Zhang, 2012]{SUNZHA12}
Sun, T. and Zhang, C.-H. (2012).
\newblock Scaled sparse linear regression.
\newblock {\em Biometrika}, 99:879--898.

\bibitem[Tibshirani, 1996]{TIB96}
Tibshirani, R. (1996).
\newblock Regression shrinkage and selection via the lasso.
\newblock {\em J. Royal. Statist. Soc B.}, 58(1):267--288.

\bibitem[Tibshirani et~al., 2016]{TIBTAYLOCTIB16}
Tibshirani, R., Taylor, J., Lockhardt, R., and Tibshiranis, R. (2016).
\newblock Exact post-selection inference for sequential regression procedures.
\newblock {\em Journal of the American Statistical Association}, 111.

\bibitem[Tukey, 1993]{TUK93B}
Tukey, J.~W. (1993).
\newblock Issues relevant to an honest account of data-based inference,
  partially in the light of {L}aurie {D}avies's paper.
\newblock Princeton University, Princeton.

\bibitem[Wasserman and Roeder, 2009]{WASSROED09}
Wasserman, L. and Roeder, K. (2009).
\newblock High dimensional variable selection.
\newblock {\em Annals of Statistics}, 37(5A):2178--2201.

\bibitem[Zhang and Zhang, 2014]{ZHAZHA14}
Zhang, C.-H. and Zhang, S. (2014).
\newblock Confidence intervals for low dimensional parameters in high
  dimensional linear models.
\newblock {\em J. R. Stat. Soc. Ser. B. Stat. Methodol.}, 76:217--242.
\end{thebibliography}

\end{document}